\documentclass{article}
\usepackage{amssymb}

\input{tcilatex}
\begin{document}

\title{Some Inflationary Einstein-Aether Cosmologies}
\author{John D. Barrow \\
DAMTP, Centre for Mathematical Sciences,\\
Cambridge University,\\
Wilberforce Rd.,\\
Cambridge CB3 0WA\\
UK\\
}
\maketitle

\begin{abstract}
We show how to derive several families of accelerating universe solutions to
an Einstein-Aether gravity theory. These solutions provide possible
descriptions of inflationary behaviour in the early universe and late-time
cosmological acceleration.

PACS numbers 04.50.Kd, 04.20.Fy, 98.80.Cq
\end{abstract}

\section{\protect\bigskip \protect\bigskip Introduction}

There has been renewed interest in Lorentz-violating theories of gravity and
their consequences for experimental gravity and cosmology. \ Donnelly and
Jacobson \cite{DJ} have provided a systematic construction of an
Einstein-Aether gravity theory of this sort that preserves locality and
covariance in the presence of an additional Lorentz-violating ('aether')
vector field. This determines a preferred rest frame at each spacetime
point, as was also considered by Gasparini \cite{gasp}, and leads to
interesting variations on the standard picture for the development of
large-scale structure in the universe that a number of authors have examined
in detail, see refs \cite{lim,li,zun,arm,zlos,meng,kob,clif}. The aether
vector field, $u_{a}$, and the metric tensor $g_{ab}$ together determine the
local spacetime structure. In an isotropic and homogeneous Friedmann
universe with expansion scale factor $a(t)$ and comoving proper time $t$,
the aether field will be aligned with the cosmic frame and is related to the
expansion rate of the universe by

\[
\nabla _{c}u^{b}=\frac{\dot{a}}{a}(g_{cb}-u_{c}u_{b}) 
\]

The Einstein equations are generalised by the contribution of an additional
stress tensor for the aether field. If the universe contains a single
self-interacting scalar field $\phi ,$ with a self interaction potential $V$
that can now be a function of $\phi $ and the expansion rate $\theta =3\dot{a%
}/a,$then the modified stress tensor of Donnelly and Jacobson \cite{DJ} is

\begin{equation}
T_{ab}=\nabla _{a}\phi \nabla _{b}\phi -(\frac{1}{2}\nabla _{c}\phi \nabla
^{c}\phi -V+\theta V_{\theta })g_{ab}  \label{tab}
\end{equation}%
This corresponds to and effective fluid with pressure $p$ and density $\rho $
of the form

\[
T_{a}^{b}\ =diag(\rho ,-p,-p,-p)
\]%
with

\begin{equation}
\rho =\frac{1}{2}\dot{\phi}^{2}+V-\theta V_{\theta }  \label{rho}
\end{equation}

\begin{equation}
p=\frac{1}{2}\dot{\phi}^{2}-V+\theta V_{\theta }-\dot{V}_{\theta }  \label{p}
\end{equation}%
with $V(\phi ,\theta )$ where $\theta =3H=3\dot{a}/a$.

The energy-momentum conservation law,

\begin{equation}
\dot{\rho}+3H(\rho +p)=0,  \label{con}
\end{equation}%
then remains as in general relativity

\begin{equation}
\ddot{\phi}+3H\dot{\phi}+V_{\phi }=0,  \label{1}
\end{equation}%
while the Friedmann equation is augmented by the contribution of the aether
stress to the energy density ($8\pi G=1=c$):

\begin{equation}
3H^{2}=\rho =\frac{1}{2}\dot{\phi}^{2}+V-\theta V_{\theta }-\frac{k}{a^{2}},
\label{2}
\end{equation}%
where $k$ is the usual Friedmann curvature parameter in the metric (in
coordinates \{$t,r,\vartheta ,\varphi $\})

\[
ds^{2}=dt^{2}-a^{2}(t)\{\frac{dr^{2}}{1-kr^{2}}+r^{2}d\vartheta
^{2}+r^{2}\sin ^{2}\vartheta d\varphi ^{2}\},
\]%
We will now set $k$ equal to zero in what follows.

The energy-momentum tensor, and the forms of the density and pressure it
contains, are reminiscent of the form required when a simple bulk viscosity
is added to a perfect fluid close to equilibrium \cite{trec, defl, wein}.
However, there are differences. The addition of a bulk viscosity $\eta (\rho
)$ to a fluid with density $\rho $ and isotropic pressure $p$ is obtained by
effecting the transformation

\[
(\rho ,p)\rightarrow (\rho ,p-\theta \eta )
\]%
in the equations (\ref{con}) and the left-hand equality in (\ref{2}), so the
Friedmann equation of general relativity ($3H^{2}=\rho -ka^{-2}$) is left
unaltered but the density conservation changes. By contrast, the
introduction of the aether field is effected by the transformation

\[
(\rho ,p)\rightarrow (\rho -\theta V_{\theta },p+\theta V_{\theta }-\dot{V}%
_{\theta }) 
\]%
which differs from the situation with bulk viscous stresses in an expanding
universe unless $V_{\theta }=0$ and $\dot{V}_{\theta }>0$.

\section{\protect\bigskip Simple exact solutions}

Solutions of these equations are of interest in two cosmological eras. The
first is in the early period where accelerated 'inflationary' expansion
might occur for a finite time interval, solving the traditional horizon,
flatness, and isotropy problems while creating a distinctive inhomogeneity
spectrum which leaves its gravitational imprint on the microwave background
radiation anisotropy and statistics. The second is at late times when the
universal expansion is observed to be accelerating because of the influence
of some gravitationally repulsive stress, aka 'dark energy'. It may be
important for a viable cosmological model to exhibit both periods of
accelerated expansion in order to be consistent with all astronomical
observations. It is not clear whether a single scalar field might be
responsible for the early and the late-time acceleration and so far there is
no compelling cosmological model in which it is.

We look for a general scale invariant solution of (\ref{1})-(\ref{2}) in
which

\begin{equation}
V(\theta ,\phi )=V_{0}\exp [-\lambda \phi ]+\sum_{r=0}^{n}a_{r}\theta
^{r}\exp [(r-2)\lambda \phi /2],  \label{3}
\end{equation}%
where $V_{0},\lambda $ and $\{a_{r}\}$ are constants. Note that the series
could be extended to negative $r$ if required.\emph{\ }

This choice of potential subsumes the simple cases with $V(\theta ,\phi
)=f(\phi )\theta ^{2}$\emph{\ }of\emph{\ }Kanno and Soda \cite{kann} and $%
V(\theta ,\phi )=f(\theta ^{2})$ of Zlosnik et al \cite{zlos} considered
earlier for specific purposes, but does not include the choice $V(\theta
,\phi )=\frac{1}{2}m^{2}\phi ^{2}+\mu \theta \phi $ explored in ref \cite{DJ}
in the context of inflationary models where $V$ exhibits a minimum in $\phi $%
.

There exist exact power-law solutions of (\ref{1})-(\ref{2}) with

\begin{eqnarray}
\phi &=&\frac{2}{\lambda }\ln t,  \label{4} \\
a &=&t^{B}, \\
\theta &=&3\frac{\dot{a}}{a}=3Bt^{-1}.  \nonumber
\end{eqnarray}%
With these choices in (\ref{3}) we have

\[
V(\theta ,\phi )=\frac{V_{0}+S_{n}}{t^{2}},
\]%
where $S_{n}$ is the finite series of constants:

\begin{equation}
S_{n}\equiv \sum_{r=0}^{n}a_{r}(3B)^{r}.  \label{5}
\end{equation}%
We see that

\[
V_{\theta }=\sum_{r=0}^{n}a_{r}\theta ^{r-1}r\exp [(r-2)\lambda \phi
/2]\equiv \frac{R_{n}}{t^{2}},
\]%
where the finite series,

\begin{equation}
R_{n}\equiv \sum_{r=0}^{n}ra_{r}(3B)^{r-1},  \label{6}
\end{equation}%
is a constant. We note also that

\[
V_{\phi }=-\lambda V_{0}\exp [-\lambda \phi ]+\sum_{r=0}^{n}a_{r}\theta
^{r}\left( \frac{r-2}{2}\right) \lambda \exp [(r-2)\lambda \phi /2]=\frac{%
\lambda }{t^{2}}(T_{n}-V_{0}), 
\]%
where the finite series

\begin{equation}
T_{n}=\sum_{r=0}^{n}a_{r}\left( \frac{r-2}{2}\right) (3B)^{r}  \label{7}
\end{equation}%
is a constant.

We see from these definitions that

\begin{equation}
S_{n}+T_{n}=\frac{3BR_{n}}{2}.  \label{8}
\end{equation}%
\ \ \ 

Substituting these expressions for $V,V_{\theta }$ and $V_{\phi }$ into the
eqns. (\ref{1})-(\ref{2}), we can obtain the algebraic constraints needed to
determine $B$ in terms of the constants $V_{0},\lambda $ and $a_{i}$ which
specify the potential completely.

From eqn. (\ref{2}) we obtain

\[
3B^{2}=\frac{2}{\lambda ^{2}}+V_{0}+S_{n}-3BR_{n},
\]%
and from eqn. (\ref{1}) we have

\[
-\frac{2}{\lambda }+\frac{6B}{\lambda }-\lambda V_{0}+\lambda T_{n}=0.
\]%
Solving these, using eqn. (\ref{8}), we find

\begin{equation}
B=\frac{2}{\lambda ^{2}}-\frac{R_{n}}{2}  \label{9}
\end{equation}%
and

\begin{equation}
V_{0}=3B^{2}-\frac{2}{\lambda ^{2}}-S_{n}+3BR_{n}=T_{n}+\frac{2}{\lambda ^{2}%
}\left( 3B-1\right) .  \label{10}
\end{equation}

We note some interesting special cases. When the $a_{i}$ are all zero, so $%
R_{n}=S_{n}=T_{n}=0,$ the potential is the familiar exponential potential 
\cite{lucc, hal, jbexp} and there is power-law inflationary solution when $%
\lambda <\sqrt{2}:$ 
\begin{eqnarray*}
a_{i} &=&0,\forall i:B=2/\lambda ^{2}\text{ }\geq \frac{1}{3}\text{\ }, \\
V_{0} &=&\frac{2}{\lambda ^{2}}\left( \frac{6}{\lambda ^{2}}-1\right) \geq 0.
\end{eqnarray*}

In general, we see that we can have power-law inflation so long as

\[
B=\frac{2}{\lambda ^{2}}-\frac{R_{n}}{2}>1.
\]%
It is instructive to look at a particular illustrative example. Suppose that
only $a_{2}$ is non-zero and the potential simplifies to

\[
V(\theta ,\phi )=V_{0}\exp [-\lambda \phi ]+a_{2}\theta ^{2}=\frac{%
V_{0}+S_{n}}{t^{2}},
\]%
then

\[
R_{n}=6Ba_{2};S_{n}=9B^{2}a_{2};T_{n}=0, 
\]%
and

\begin{eqnarray*}
B &=&\frac{2}{\lambda ^{2}(1+3a_{2})}, \\
V_{0} &=&\ \frac{2}{\lambda ^{2}}\left( 3B-1\right) =B\left( 3B-1\right)
(1+3a_{2}).
\end{eqnarray*}%
Here, we see explicitly the requirement on the coupling parameter $a_{2}$
for inflation to occur. We note that it is possible for $a_{2}\neq 0$ to
create inflationary expansion (ie $B>1$) in cases where the same value of $%
\lambda $ would not lead to inflation when the aether field is absent.
Similar properties are shared by the general case when all the $a_{i}$ are
non-zero.

\section{ Further exact solutions}

The general system of equations we have solved also simplifies in ways that
permit phase portraits to be created if required. If we differentiate the
Friedmann equation we get

\bigskip

\begin{eqnarray*}
6H\dot{H} &=&\dot{\phi}\ddot{\phi}+V_{\phi }\dot{\phi}+V_{\theta }\dot{\theta%
}-V_{\theta }\dot{\theta}-\theta \frac{d}{dt}(V_{\theta }) \\
6H\dot{H} &=&\dot{\phi}\ddot{\phi}+V_{\phi }\dot{\phi}-\theta \left(
V_{\theta \phi }\dot{\phi}+V_{\theta \theta }\dot{\theta}\right) \\
2\dot{H} &=&-\dot{\phi}^{2}-\dot{\phi}V_{\theta \phi }-9H\dot{H}V_{\theta
\theta }
\end{eqnarray*}%
and

\begin{eqnarray*}
3H^{2} &=&\frac{1}{2}\dot{\phi}^{2}+V-\theta V_{\theta } \\
\ddot{\phi}+3H\dot{\phi}+V_{\phi } &=&0
\end{eqnarray*}

In order to solve the last three equations, suppose $V$ has a general
separable form:

\[
V(\theta ,\phi )=U(\phi )+\mu f(\theta )g(\phi ) 
\]%
then

\begin{eqnarray*}
3H^{2} &=&\frac{1}{2}\dot{\phi}^{2}+U+\mu g(f-\theta f_{\theta }) \\
\ddot{\phi}+3H\dot{\phi}+U_{\phi }+\mu fg_{\phi } &=&0
\end{eqnarray*}%
There is a family of special solutions for which

\begin{equation}
f-\theta f_{\theta }=C  \label{s1}
\end{equation}%
and so

\begin{equation}
f=C+F\theta  \label{s2}
\end{equation}%
with $C,F$ constants.

The case considered by Donnelly and Jacobson in \cite{DJ} is

\[
U=\frac{1}{2}m^{2}\phi ^{2}
\]%
\begin{eqnarray*}
f &=&M\theta  \\
g &=&\phi .
\end{eqnarray*}%
In the case where (\ref{s1}) holds, we can choose

\begin{equation}
U(\phi )=U_{0}\exp [-\lambda \phi ]  \label{exp}
\end{equation}%
so

\begin{eqnarray*}
3H^{2} &=&\frac{1}{2}\dot{\phi}^{2}+U_{0}\exp [-\lambda \phi ] \\
\ddot{\phi}+3H\dot{\phi}-\lambda U_{0}\exp [-\lambda \phi ]+\mu (C+F\theta
)g_{\phi } &=&0 \\
2\dot{H} &=&-\dot{\phi}^{2}-\dot{\phi}V_{\theta \phi }-9H\dot{H}V_{\theta
\theta }=-\dot{\phi}^{2}-\dot{\phi}\mu g_{\phi }f_{\theta }-9H\dot{H}\mu
gf_{\theta \theta }
\end{eqnarray*}%
But in our special case $f_{\theta \theta }=0$ and $f_{\theta \ \ \ }=F,$ so

\[
2\dot{H}=\dot{\phi}^{2}-\dot{\phi}\mu g_{\phi }F
\]%
The recipe for solving this system is to pick $g(\phi )$ then $\phi (t)$;
solve for $H(t)$ and hence use $t(\phi )$ to obtain $H(\phi )$ and find the
constraint on the constants from the Friedmann equation; see for example 
\cite{jdb1,jdb2,jdb3} for corresponding results using this method in general
relativistic cosmologies.

Example 1: Pick 
\begin{eqnarray*}
f &=&M\theta \\
g &=&\phi \\
\phi &=&A\ln [\tanh (\lambda t)]
\end{eqnarray*}%
so

\[
\exp [\phi /A]=\tanh (\lambda t) 
\]

\[
\dot{\phi}=2A\lambda \func{cosech}(2\lambda t) 
\]

\[
2H=2H_{0}-\mu \phi -\int \dot{\phi}^{2}dt=2H_{0}-\mu \phi +2\lambda
A^{2}\cosh (\phi /A)
\]%
Therefore, (if $H_{0}=0$), we have

\[
U(\phi )=\frac{3\mu ^{2}\phi ^{2}}{4}-3\lambda \mu A^{2}\phi \cosh (\phi
/A)+\lambda ^{2}A^{2}\left[ (3A^{2}-2)\cosh ^{2}(\phi /A)+2\right] 
\]%
Or, asymptotically, keeping $H_{0}\neq 0,$ as $t\rightarrow \infty $

\[
H\rightarrow H_{0}\ \ +\lambda A^{2} 
\]%
and $\mu $ has no effect.

Example 2

\begin{eqnarray*}
\phi &=&A\func{cosech}(\lambda t) \\
H &=&H_{0}-\frac{A\mu }{2\sinh (\lambda t)}+\frac{\lambda A^{2}}{6}\coth
^{3}(\lambda t) \\
a &=&a_{0}e^{H_{0}t}\left[ \tanh (\lambda t/2)\right] ^{-\frac{A\mu }{%
2\lambda }}\left[ \sinh (\lambda t)\right] ^{\frac{A^{2}}{6}}\exp [-\frac{%
A^{2}}{12}\coth ^{2}(\lambda t)]
\end{eqnarray*}%
and as $t\rightarrow \infty $ we have

\[
a\rightarrow a_{0}\exp [H_{0}t+\frac{\lambda A^{2}t}{6}] 
\]%
Again, $\mu $ has no effect on the inflation.

\bigskip Example 3

If we change the time variable to $\tau $, where

\[
d/dt=V^{1/2}d/d\tau ,
\]

\bigskip and denote $d/d\tau $ by $^{\prime }$, and put

\[
a(t)=exp[\alpha (\tau )]
\]%
then 

\begin{eqnarray*}
\phi ^{\prime \prime }+\frac{V_{\phi }}{2V}\phi ^{\prime 2}+\frac{V_{\phi }}{%
V} &=&0 \\
3\alpha ^{\prime 2} &=&\frac{1}{2}\phi ^{\prime 2}+1-\frac{3V_{\theta }}{%
2V^{1/2}}\alpha ^{\prime }
\end{eqnarray*}%
and we have an autonomous system when

\begin{equation}
\frac{V_{\phi }}{V}=-\lambda =\text{ constant}  \label{c1}
\end{equation}

and

\begin{equation}
\frac{V_{\theta }}{V^{1/2}}=\mu =\text{ constant}  \label{c2}
\end{equation}%
This system can be explored by standard phase plane techniques although we
shall not do that here. These constraints (\ref{c1})-(\ref{c2}) are
satisfied by the choice

\[
V(\theta ,\phi )=\frac{\mu ^{2}}{4}(\theta +\theta _{0})^{2}+V_{0}\exp
[-\lambda \phi ]. 
\]%
Setting $\theta _{0}=0$ for simplicity, we see there is a particular exact
solution in $t$ time:

\begin{eqnarray*}
a(t) &\varpropto &t^{2/\lambda ^{2}(1+3\mu )} \\
\phi  &=&\frac{2}{\lambda }\ln (t) \\
V_{0} &=&\frac{2[6-\lambda ^{2}(1+3\mu )]}{\lambda ^{4}(1+3\mu )}
\end{eqnarray*}%
which reduces to the familiar power-law inflation model for an exponential
potential when $\mu =0$. However, when $\mu \neq 0$ we see that the aether
field has a strong effect and enables inflation to occur in situations ($%
\lambda ^{2}>2$) where it is impossible in the absence of the aether field.
The introduction of $\theta _{0}\neq 0$ for a combination of power-law and
exponential expansion.

\section{\protect\bigskip Discussion}

We have shown how a series of simple ans\"{a}tze allow exact solutions to be
found for Einstein-Aether cosmologies. These models provide exact
descriptions of inflationary dynamics in the very early universe or the
transition to accelerated expansion at recent cosmological epochs. They show
the explicit contribution of the aether field to creating accelerated
expansion in situations where inflation would not occur in its absence. They
can also be extended to include simple 'tracker' solutions and allow
Einstein-Aether theories to be more closely tested by cosmological data
sets.

\end{document}